\documentclass{appolb}
\usepackage{epsfig}
\usepackage{amssymb}

\begin{document}

\title{Extinction statistics in N random interacting species
\thanks{Presented at the $19^{th}$ Marian Smoluchowski Symposium on Statistical Physics,
Krak\'ow, Poland, May 14-17, 2006}
}

\author{Alessandro Fiasconaro$^{a,b,c}$\footnote{e-mail: afiasconaro@gip.dft.unipa.it}
and Bernardo Spagnolo$^a$
\address{$^a$Dipartimento di Fisica e Tecnologie Relative, Universit\`a di
Palermo\\
CNISM - Unit\`a di Palermo, Group of Interdisciplinary
Physics\footnote{http://gip.dft.unipa.it} \\
Viale delle Scienze, I-90128 Palermo, Italy\\
$^b$Marian Smoluchowski Institute of Physics, Jagellonian University\\
Reymonta 4, 30–059 Kraków, Poland\\
$^c$Mark Kac Center for Complex Systems Research, Jagellonian University\\
Reymonta 4, 30–059 Kraków, Poland}}

\maketitle

\date{\today}

\begin{abstract}
A randomly interacting N-species Lotka-Volterra system in the
presence of a Gaussian multiplicative noise is analyzed. The
investigation is focused on the role of this external noise into the
statistical properties of the extinction times of the populations.
The distributions show a Gaussian shape for each noise intensity
value investigated. A monotonic behavior of the mean extinction time
as a function of the noise intensity is found, while a nonmonotonic
behavior of the width of the extinction time probability
distribution characterizes the dynamical evolution.
\end{abstract}



\section{Introduction}
Generalized Lotka-Volterra equations have been used in recent years
to describe the dynamics of various kind of population species,
which are main components in complex
ecosystems~\cite{Mobilia}-\cite{Katja04}. To understand the complex
behavior of such ecosystems is crucial to analyze the role played by
the external noise on the dynamics. It has become increasingly
evident that nonlinearity and noise play an important role in such
complex dynamics. Recently, in fact, noise-induced effects in
population dynamics, such as pattern formation, stochastic
resonance, noise delayed extinction, quasi periodic oscillations
etc... have been investigated with increasing
interest~\cite{Soc01}-\cite{Rus00}. Complex ecological systems
evolve towards the equilibrium states through the slow process of
nonlinear relaxation, which is strongly dependent on the random
interaction between the species, the initial conditions and the
random interaction with environment. One of the open problems of
such ecosystems is the investigation of the time scales of
extinction and survival of the species and their related statistics.
Various factors affecting extinction such as migration, chaos,
interaction between species, spatial synchronization, etc., have
been discussed in the literature~\cite{Amritkar}-\cite{Vandermeer}.
However, there is lack of investigation on the role of external
noise on the extinction process, which is the main focus of this
paper. The mathematical model used to analyze the dynamics of $N$
biological species, with spatially homogeneous densities, is the
generalized Lotka-Volterra system. We consider a Malthus-Verhulst
modelling of the self regulation mechanism and an external
multiplicative noise source, taking the environment interaction into
account~\cite{Ciu96,Spa02}. Within this model we analyzed the role
of the noise in the statistical properties of the extinction times
of the populations. Specifically a monotonic behavior of the mean
extinction time as a function of the noise intensity is observed.
The width of the distribution of the extinction times, however, has
a nonmonotonic behavior as a function of the noise intensity.

\section{The model}

The dynamical evolution of our ecosystem composed by $N$ interacting
species in a noisy environment (climate, disease, etc...) is
described by the following generalized Lotka-Volterra equations with
a multiplicative noise, in the framework of Ito stochastic
calculus~\cite{Gardiner}

\begin{equation}
d n_i(t) = \left[ \left(\left(\alpha + \frac{\epsilon}{2} \right) -
n_i(t) + \sum_{j\neq i} J_{ij}n_j(t) \right)dt + \sqrt{\epsilon}
dw_i\right] n_i(t)\mbox{,}
\label{langevin}
\end{equation}
where \textit{$n_i(t) \geq 0$} is the population density of the
\textit{$i^{th}$} species at time \textit{$t$} and $i = 1,...,N$. In
Eq.~(\ref{langevin}), the first two terms describe the development
of the \textit{$i^{th}$} species without interacting with other
species, \textit{$\alpha$} is the growth parameter, and
\textit{$J_{ij}$} is the interaction matrix, which models the
interaction between different species ($i\neq j$). Here
\textit{$w_i$} is the Wiener process whose increment \textit{$dw_i$}
satisfy the usual statistical properties \emph{$ \langle dw_i(t)
\rangle \thinspace = \thinspace 0$}, and \emph{$\langle
dw_i(t)dw_j(t^{\prime})\rangle \thinspace = \thinspace
\delta_{ij}\delta(t-t^{\prime}) dt$}. The interaction matrix
\emph{$J_{ij}$} has elements randomly distributed according to a
Gaussian distribution with \emph{$\langle J_{ij}\rangle = 0$},
\emph{$\langle J_{ij} J_{ji}\rangle = 0$}, and \emph{$\sigma^2_j =
J^2/N$}. Our ecosystem contains, therefore, 50$\%$ of prey-predator
interactions (\emph{$J_{ij}< 0$} and \emph{$J_{ji}
> 0$}), 25$\%$ competitive interactions (\emph{$J_{ij}>0$} and
\emph{$J_{ji}>0$}), and 25$\%$ symbiotic interactions
(\emph{$J_{ij}<0$} and \emph{$J_{ji}<0$}). We consider all species
equivalent so that the characteristic parameters of the ecosystem
are independent of the species. The formal solution of
Eq.~(\ref{langevin}) is

\begin{equation} n_i(t) = \frac{n_i(0) exp\left[\alpha t +\sqrt{\epsilon} w_i(t) +
\int_{0}^{t} dt^{\prime}\sum_{j\neq i}J_{ij}n_j(t^{\prime}))\right]}
{1+ n_i(0) \int_{0}^{t}dt^{\prime}  exp\left[\alpha t^{\prime}
+\sqrt{\epsilon} w_i(t^{\prime}) + \int_{0}^{t^{\prime}}
dt^{''}\sum_{j\neq i}J_{ij}n_j(t^{''}))\right]}\;,
\label{sol-langevin}
\end{equation}
where the term $h_i(t)=\sum_{j\neq i}J_{ij}n_j(t)$ represents the
influence of other species on the differential growth rate of the
\textit{$i^{th}$} population. The dynamical behavior of the
$i^{th}$ population depends on the time integral of the term
$h_i(t)$ and the time integral in the denominator of
Eq.~(\ref{sol-langevin}). By considering the deterministic
dynamics (in the absence of external noise ($\epsilon = 0$)), with
a large number of interacting species (that is large interaction
random matrix), we can assume that the term $h_i(t)$ is Gaussian
with zero mean and variance $\sigma_{h_i}^2 = \Sigma_{j,k} \langle
J_{ij} J_{ik}\rangle \langle n_j n_k\rangle = J^2 \langle
n_i^2\rangle$, with $\langle J_{ij} J_{ik}\rangle =
\delta_{jk}\frac{J^2}{N}$. In the absence of external noise, from
the fixed-point equation $n_i(\alpha - n_i + h_i) = 0$, the
stationary probability distribution of the populations is the sum
of a truncated Gaussian distribution at $n_i =0$ ($n_i > 0$) and a
delta function for extinct species. The initial values of the
populations $n_i(0)$ have also Gaussian distribution with mean
value $\langle n_i(0)\rangle = 1$, and variance $\sigma^2_{n(0)} =
0.03$. The interaction strength $J$ determines two different
dynamical behaviors of the ecosystem. Above a critical value
\emph{$J_c = 1.1$}, the system is unstable and at least one of the
populations diverges. Below $J_c$ the system is stable and
asymptotically reaches an equilibrium state. The equilibrium
values of the populations depend both on their initial values and
on the interaction matrix. If we consider a quenched random
interaction matrix, the ecosystem has a great number of
equilibrium configurations, each one with its attraction basin.
For an interaction strength $J = 1$ and an intrinsic growth
parameter $\alpha = 1$ we obtain: $\langle n_i \rangle = 1.4387,
\langle n^{2}_i \rangle = 4.514,$ and $\sigma^{2}_{n_i} = 2.44$.
These values agree with that obtained from numerical simulation of
Eq.~(\ref{langevin}).

The statistics of the species extinction has been analyzed using
the mean extinction time $\langle t_m \rangle$, defined as

\begin{equation}
  \langle t_m \rangle = \frac{1}{N_{exp}}\sum_{i=1}^{N_{exp}}t_{m},
\label{met}
\end{equation}
and its variance

\begin{equation}
  \sigma^2 =  \langle t_m^2 \rangle -  \langle t_m \rangle^2  .
\label{sigma}
\end{equation}
Here $\langle t_m \rangle$ is an ensemble average ($N_{exp}$ is
the number of simulative experiments), $t_m$ is the average
extinction time over the number of populations $N$

\begin{equation}
  t_m = \frac{1}{N} \sum_{i=1}^{N} t_{i,m},
\label{tm}
\end{equation}
and $t_{i,m}$ is the extinction of the $i$-th population in the
$m$-th experiment.

\section{Results and Comments}

 The parameters used in our simulation are:
 $\alpha=1.2$, $J=1$, $\sigma_J^2= 0.005$, $N = 400$. The
number of simulative experiments is $N_{exp}=1000$, and the initial
values of the average population and its standard deviation are:
$\langle n_i \rangle = 1$, $\sigma_{n_o}^2 = 0.03$. The dynamics of
various species are different even if they are equivalent according
to the parameters in the dynamical equation~(\ref{langevin}).
However to change the species index by fixing the random matrix or
to change the random matrix by fixing the species index should be
equivalent as regards the asymptotic dynamical regime.

\begin{figure}[htb]
 \centering
 \includegraphics[width=12.5 cm]{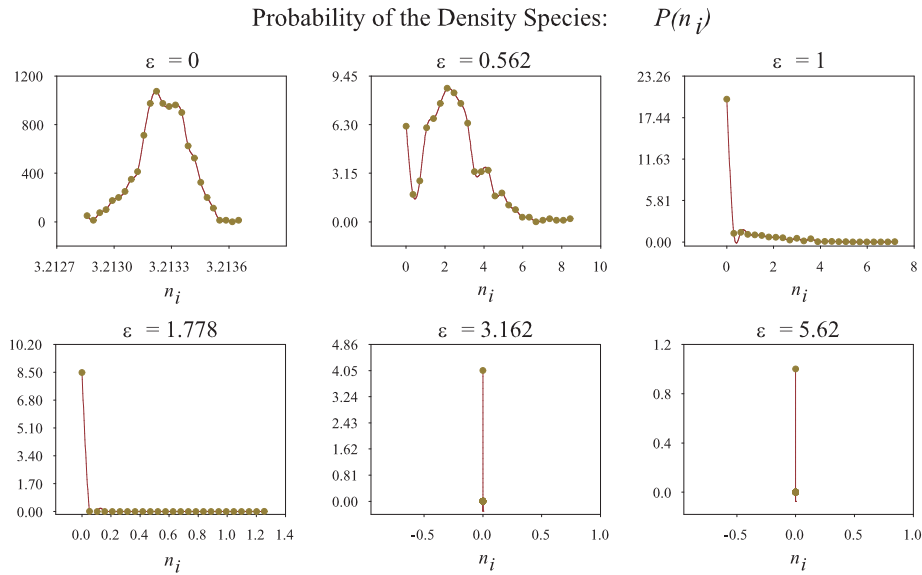}
\caption{Long time probability distribution of the species densities
for different external noise intensities $\epsilon.$ Namely
$\epsilon = 0, 0.562, 1, 1.778, 3.162, 5.62$. Around the value
$\epsilon = 0.562$ the distribution becomes asymmetric, and for
$\epsilon > 1.778$, all the species are extinct.}
 \label{density}
\end{figure}

In the presence of external noise ($\epsilon \neq 0$) we calculate
the long time probability distribution of the species density for
different values of the noise intensity. These are shown in
Fig.~\ref{density}.

\begin{figure}[htb]
 \centering
 \includegraphics[width=8 cm]{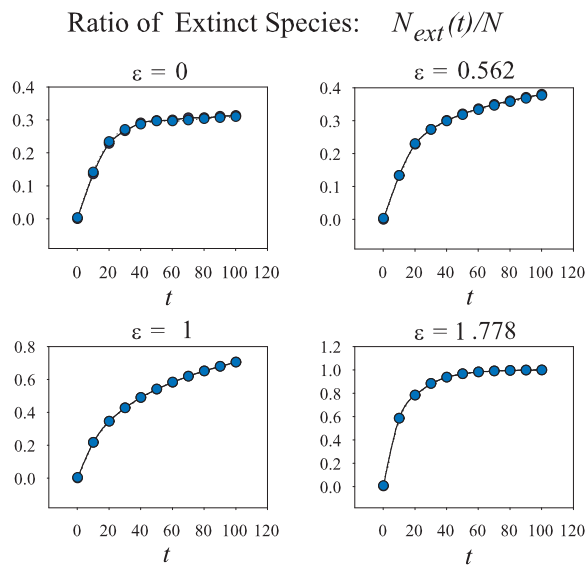}
\caption{Time evolution of the normalized number of the extinct
species for different noise intensities $\epsilon$. Namely:
$\epsilon = 0, 0.562, 1, 1.778$.}
 \label{extinct}
\end{figure}

For increasing external noise intensity we obtain a larger
probability distribution with a lower maximum (see the different
scales in Figs.~\ref{density} for different noise intensity values).
The distribution is asymmetric for $\epsilon = 0.562$ and tends to
become a truncated delta function around the zero value ($P(n_i) =
\delta(n_i)$ for $n_i > 0$, and $P(n_i) = 0$ for $n_i \leq 0$), for
further increasing noise intensity. Specifically for high values of
noise intensity (namely for $\epsilon
> 1.778$) we strongly perturb the population dynamics and because of
the presence of an absorbing barrier at $n_i = 0$~\cite{Ciu96}, we
obtain quickly the extinction of all the species. To confirm this
picture we calculate the time evolution of the average number of
extinct species for different noise intensities. This time behavior
is shown in Fig.~\ref{extinct}. We see that this number increases
with noise intensity, obtaining a rapid transient dynamics of the
system towards the extinction final state for $\epsilon \geq 1.778$.
This means that the species rapidly die and the probability
distribution of the species density confines accordingly.

In the following Fig.~\ref{pt} we show the probability distribution
function (PDF) of the extinction times of the species. The shape of
the distribution is Gaussian in the deterministic regime ($\epsilon
= 0$) and in the presence of the external noise ($\epsilon \neq 0$).
For low noise intensities the probability distribution becomes
larger and lower until reaches the value of $\epsilon = 1$. After
this value of noise intensity the distribution becomes narrow and
higher. The mean extinction time, which is easily visible from the
figure because of the Gaussian shape distribution, decreases
monotonically with increasing noise intensity. In this figure it is
shown a well defined extinction time windows of the species, moving
towards the absorbing barrier at $n_i = 0$, with increasing noise
intensity.
\begin{figure}[htb]
 \centering
 \includegraphics[width=12.5 cm]{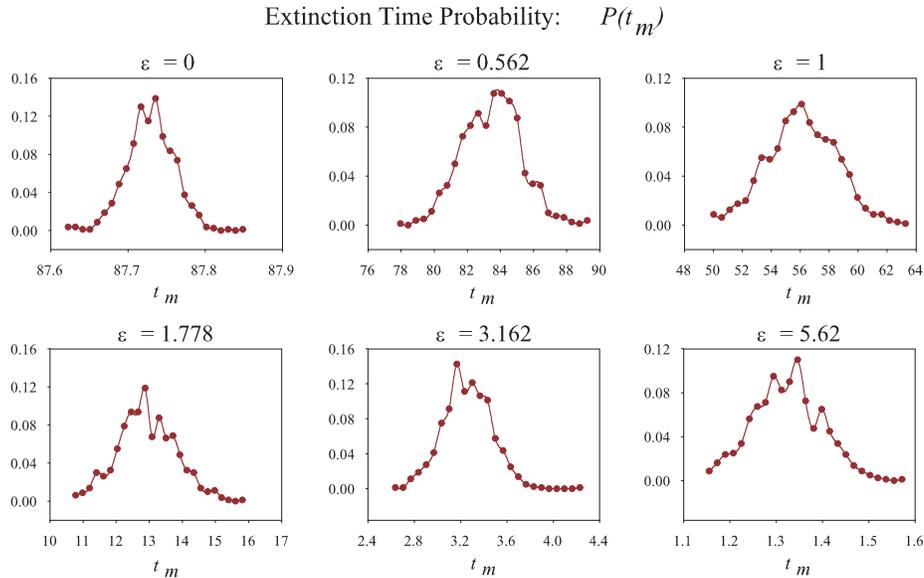}
\caption{Probability distribution function of the extinction times
of the species, for different values of the noise intensity. Namely:
$\epsilon = 0, 0.562, 1, 1.778, 3.162, 5.62$. All the PDFs show a
Gaussian shape distribution.}
 \label{pt}
\end{figure}

This behavior is due to the presence of the external noise and the
absorbing barrier. In fact, in the deterministic case ($\epsilon =
0$), the Gaussian distribution of the extinction times is due only
to the random interaction matrix. The characteristic values of the
distribution, that is the mean and the variance, depend on the
choice of the parameters of the model, that is the growth
parameter $\alpha$, the interaction strength $J$ and the initial
conditions. A small amount of noise forces the system to sample
more of the available range in the parameter space and therefore
moves lightly the system towards the extinction. The average
extinction time at $\epsilon = 0.562$ is less than that in the
absence of external noise. This enlargement and lowering of the
PDF continues until the noise intensity reaches the value of the
interaction strength $J = 1$. After that the external noise
prevails on the interaction matrix term and the extinction process
proceeds quickly because of the presence of the absorbing barrier
at $n_i = 0$ (see Eq.~(\ref{langevin})). Almost all the species
extinguish in short times around a very low mean extinction time.
At $\epsilon = 3.162$, for example, $\langle t_m \rangle \sim 3.3
$. Increasing the noise intensity ($\epsilon > 1$), therefore, the
PDF becomes narrower and higher.

As can be seen in Fig.~\ref{pt} for $\epsilon=3.162$ and
$\epsilon=5.62$ the species extinction happens in few time units, so
that the probability of density species vanishes for the same values
(see Fig.~\ref{density}).

This peculiar behavior of the PDF of extinction times gives rise to
the nonmonotonic behavior of the variance of the same quantity as a
function of the noise intensity. This is shown in the following
Fig.~\ref{noise}.
\begin{figure}[htb]
 \centering
 \includegraphics[width=9 cm]{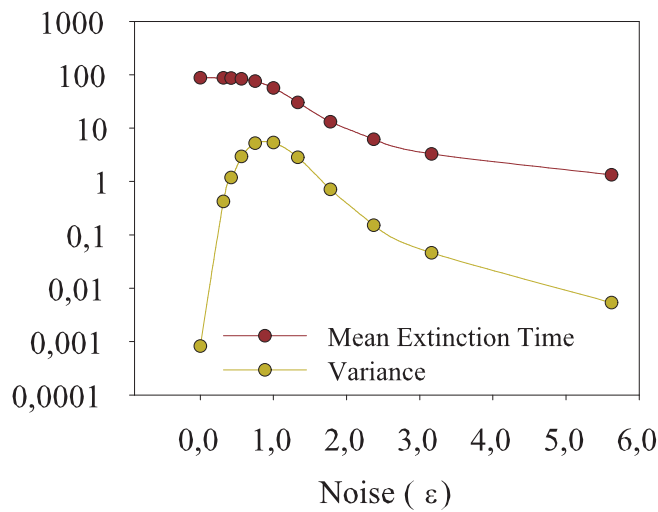}
\caption{Mean extinction time and variance as a function of the
noise intensity $\epsilon$. The variance shows a nonmonotonic
behavior with a maximum at $\epsilon \simeq 1$ and very low values
at higher noise intensities.}
 \label{noise}
\end{figure}

From this figure, the maximum of the variance at the noise
intensity $\epsilon \simeq 1$ and the very small values of the
variance at high noise intensities, are clearly visible. In the
same figure the monotonic behavior of the mean extinction time
$\langle t_m \rangle$ is shown. Calculation have been repeated for
different number of populations, namely: $N=100, 200, 300, 400$.
In all the calculations the qualitative behaviors of the mean
extinction time and its variance are the same than those reported
in Fig~\ref{noise}.

We didn't reveal any power law decay for the probability
distribution of extinction times as found in previous
investigations~\cite{Pigolotti}, but we plan to make a more detailed
investigation of species lifetime distribution in a forthcoming
paper.

\section{Conclusions}

The analysis of the dynamics of  ecosystem composed by \emph{N}
random interacting species has been performed in the presence of
multiplicative noise. The probability density of the extinction time
of the species ($P(t)$) has been evaluated for various noise
intensities. The extinction times $t_m$s are Gaussian distributed
with a mean value monotonically decreasing as a function of the
noise intensity. The variance of the extinction times shows a
nonmonotonic behavior, which characterizes the transient dynamics of
the $N$ random interacting species model.

\section*{Acknowledgements}
This work was supported by MIUR, CNISM and INFM-CNR.


\begin{thebibliography}{99}

\bibitem{Mobilia}
M. Mobilia, I. T. Georgiev, and U. C. T$\ddot{a}$uber, Phys Rev. E
\textbf{73}, 040903 (2006); A. Venaille, P. Varona, and M. I.
Rabinovich, Phys Rev. E \textbf{71}, 061909 (2005).

\bibitem{Mankin}
A. Sauga and R. Mankin, Phys Rev. E \textbf{71}, 062103 (2005).

\bibitem{McKane} A. J. McKane and T. J. Newman, Phys. Rev. Lett. \textbf{94},
218102 (2005); A. Shabunin, A. Efimov, G.A. Tsekouras, et al.,
Physica A \textbf{347}, 117-136 (2005).

\bibitem{Tokita} K. Tokita, Phys. Rev. Lett. \textbf{93}, 178102 (2004); G. J.
Ackland and I. D. Gallagher, Phys. Rev. Lett. \textbf{93}, 158701
(2004); R. Mankin, A. Sauga, A. Ainsaar, et al., Phys. Rev. E
\textbf{69}, 061106 (2004); Y. De Decker, G.A. Tsekouras, A.
Provata, et al., Phys. Rev. E \textbf{69}, 036203 (2004); G.A.
Tsekouras, A. Provata, C. Tsallis, Phys. Rev. E \textbf{69}, 016120
(2004); A. Provata, G.A. Tsekouras, F. Diakonos, et al., Fluct.
Noise Lett. \textbf{3}, L241-L250 (2003).

\bibitem{Mur02}
J. D. Murray, \emph{Mathematical Biology I} (Springer, Berlin,
2002).

\bibitem{Katja04}
C. Escudero, J. Buceta, F. J. de la Rubia, and Katja Lindenberg,
Phys. Rev. E \textbf{69}, 021908 (2004); T. J. Kawecki and R. D.
Holt, Am. Nat. \textbf{160}, 333 (2002); Michel Droz and Andrzej
P\c{e}kalski, Phys. Rev. E \textbf{69}, 051912 (2004).

\bibitem{Giardina}
I. Giardina, J. P. Bouchaud, M. Mezard, J. Phys. A: Math. Gen.
\textbf{34}, L245 (2001); H Rieger, J. Phys. A: Math. Gen.
\textbf{22}, 3447 (1989).

\bibitem{Soc01}
J. E. S. Socolor, S. Richards, and W. G. Wilson, Phys. Rev. E
\textbf{63}, 041908 (2001).

\bibitem{Fia04}
A. Fiasconaro, D. Valenti and B. Spagnolo, Acta Phys. Pol. B
\textbf{35}, 1491 (2004); D. Valenti, A. Fiasconaro and B.
Spagnolo, Acta Phys. Pol. B \textbf{35}, 1481 (2004); A. La
Barbera and B. Spagnolo, Physica A \textbf{314}, 120 (2001).

\bibitem{Spa03}
B. Spagnolo A. Fiasconaro and D. Valenti, Fluct. Noise Lett.
\textbf{3}, L177 (2003); B. Spagnolo and A. La Barbera, Physica A
\textbf{315}, 201 (2002); A. F. Rozenfeld Rozenfeld, C.J. Tessone,
E. Albano, H.S. Wio, Phys. Lett. A \textbf{280}, 45 (2001).

\bibitem{Sci99}
See the special section on \emph{"Complex Systems"}, Science
\textbf{284}, 79-107 (1999); the special section on \emph{"Ecology
through Time"}, Science \textbf{293}, 623-657 (2001).

\bibitem{Val04}
D. Valenti, A. Fiasconaro and B. Spagnolo, Physica A \textbf{331},
477 (2004).

\bibitem{Spa04}
B. Spagnolo D. Valenti, A. Fiasconaro, Math. \ Biosciences \ and
Eng. \textbf{1}, 185 (2004).

\bibitem{Rus00}
D. F. Russel, L. AQ. Wilkens and F. Moss, Nature \textbf{402}, 291
(2000).

\bibitem{Amritkar}
R. E. Amritkar and G. Rangarajan, Phys. Rev. Lett. \textbf{96},
258102 (2006).

\bibitem{Pigolotti}
S. Pigolotti, A. Flammini, M. Marsili, and A. Maritain, Proc. Natl.
Acad. Sci. USA \textbf{102}, 15747 (2005)

\bibitem{Coppex}
F. Coppex, M. Droz, and A. Lipowski, Phys. Rev. E \textbf{69},
061901 (2004).

\bibitem{Vandermeer}
J. Vandermeer et al., Proc. Natl. Acad. Sci. USA \textbf{99}, 8731
(2002); G. Abramson and D. H. Zanette, Phys. Rev. E \textbf{57},
4572 (1998); M. Heino, V. Kaitala, E. Ranta, and J. Lindstrom, Proc.
R. Soc. B \textbf{264}, 481 (1997); D. J. Earn, P. Rohani, and B.
Grenfell, Proc. R. Soc. B \textbf{265}, 7 (1998).

\bibitem{Ciu96}
S. Ciuchi, F. de Pasquale and B. Spagnolo, Phys. Rev. E
\textbf{54}, 706 (1996); ibid. \textbf{47}, 3915 (1993); P.
Barrera, S. Ciuchi and B. Spagnolo, J. Phys. A: Math. Gen.
\textbf{26}, L559-L565 (1993).

\bibitem{Spa02}
B. Spagnolo, M. A. Cirone, A. La Barbera and F. de Pasquale, J.
Phys.: Condens. Matter \textbf{14}, 2247 (2002); M. A. Cirone, F. de
Pasquale and B. Spagnolo, Fractals \textbf{11}, 217 (2003); B.
Spagnolo, D. Valenti and A. Fiasconaro, Prog. Theor. Phys. Suppl.
\textbf{157}, 312-316 (2005); A. Fiasconaro, D. Valenti and B.
Spagnolo, Eur. J. Phys. B \textbf{50}, 1-2, 189 (2006).

\bibitem{Gardiner}
C.~W.~Gardiner {\em Handbook of stochastic methods for physics,
chemistry and the natural sciences}, (Springer, Berlin, 1993).

\end{thebibliography}
\end{document}